\documentclass[twocolumn,aps,prl,nofootinbib,groupedaddress,amsmath,amssymb,longbibliography]{revtex4-1}
\usepackage{graphicx}
\usepackage{bm}
\usepackage[colorlinks=true,	linkcolor=blue,urlcolor=blue,anchorcolor=blue,citecolor=blue,bookmarksnumbered]{hyperref}
\usepackage{float}

\usepackage{changes}
\usepackage{braket}

\newcommand{\nR}{n_{\rm R}}
\newcommand{\gammac}{\gamma_{\rm C}}
\newcommand{\gammar}{\gamma_{\rm R}}
\newcommand{\gr}{g_{\rm R}^{\rm 1D}}
\newcommand{\gc}{g_{\rm C}^{\rm 1D}}
\newcommand{\R}{R^{\rm 1D}}

\begin{document}

\title{Controllable Bistability and Squeezing of Confined Polariton Dark Solitons}
\author { Gang Wang$^{1}$, Kexin Wu$^{1}$, Yang Liu$^{1}$, Weibin Li$^{2}$,  and Yan Xue$^{1,*}$}
\affiliation{$^{1}$College of Physics, Jilin University, Changchun 130012, P. R. China}
\affiliation{$^{2}$School of Physics and Astronomy, and Centre for the Mathematics and Theoretical Physics of Quantum Non-Equilibrium Systems, University of Nottingham, Nottingham, NG7 2RD, United Kingdom}
\affiliation{email: $^{*}$ xy410@jlu.edu.cn}

\begin{abstract}
The generation of squeezed light in semiconductor materials opens opportunities for building on-chip devices that are operated at the quantum level. Here we study theoretically a squeezed light source of polariton dark solitons confined in a geometric potential well of semiconductor microcavities in the strong coupling regime. We show that polariton dark solitons of odd and even parities can be created by tuning the potential depth. When driving the potential depth linearly, a bistability of solitons with the two different parities can be induced. Strong intensity squeezing is obtained near the turning point of the bistability due to the large nonlinear interaction, which can be controlled by Feshbach resonance. The phase diagram of the bistability and squeezing of the dark solitons is obtained through large scale numerical calculations. Our study contributes to the current efforts in realizing topological excitations and squeezed light sources with solid-state devices.  
\end{abstract}


\maketitle

\textit{\textbf{Introduction.}---}The peculiar light-matter composition of polaritons \cite{Sanvitto,Kavokin} allows the formation of Bose-Einstein condensation (BEC) \cite{Kasprzak_2006,Amo_2009,Xiong_2020,Schneider_2020} at elevated temperatures. The two-body interaction produced by the matter component opens a pathway to explore a broad variety of nonlinear phenomena in the solid-system setting \cite{2019natphon_Lagoudakis,2009prl_Shelykh,2015prl_Flach}, where the nonlinearity is tunable either by modulating the pump light power~\cite{Carusotto_2013,Valle_2020} or through the Feshbach resonance technology \cite{2019prl_Feshbach,2014nphy_Deveaud}. In particular, topological excitation of dark solitons attracts great interest rooted from the fundamental importance and all-optical signal processing applications \cite{Blair_1999}. Due to the giant  nonlinearity available in polariton condensates  dark solitons can form at the sub-millimeter length scales \cite{Amo_2011} and on picoseconds time scales \cite{2017prl_Ma}. Inspired by the high controllability  and on-chip integration of the semiconductor microcavity, enormous efforts have been spent on the study of nonlinear phenomena in the semi-classical regime, including dark \cite{2014prl_Yan,Walker_2017} and half-dark \cite{Hivet_2012,2015_spinor-half-vortex_sanvitto} solitons, dark soliton trains \cite{2014prl_Flayac} and their bistability \cite{2016Bloch_solitonbistability}, Cherenkov radiation of solitons \cite{2017Nc_Cherenkov_Skryabin}, dark-soliton molecules \cite{Anne2020,2020prr_Bramati}, etc.. 
\begin{figure} [!htb]
	\centering
	\includegraphics[width=0.9\linewidth]{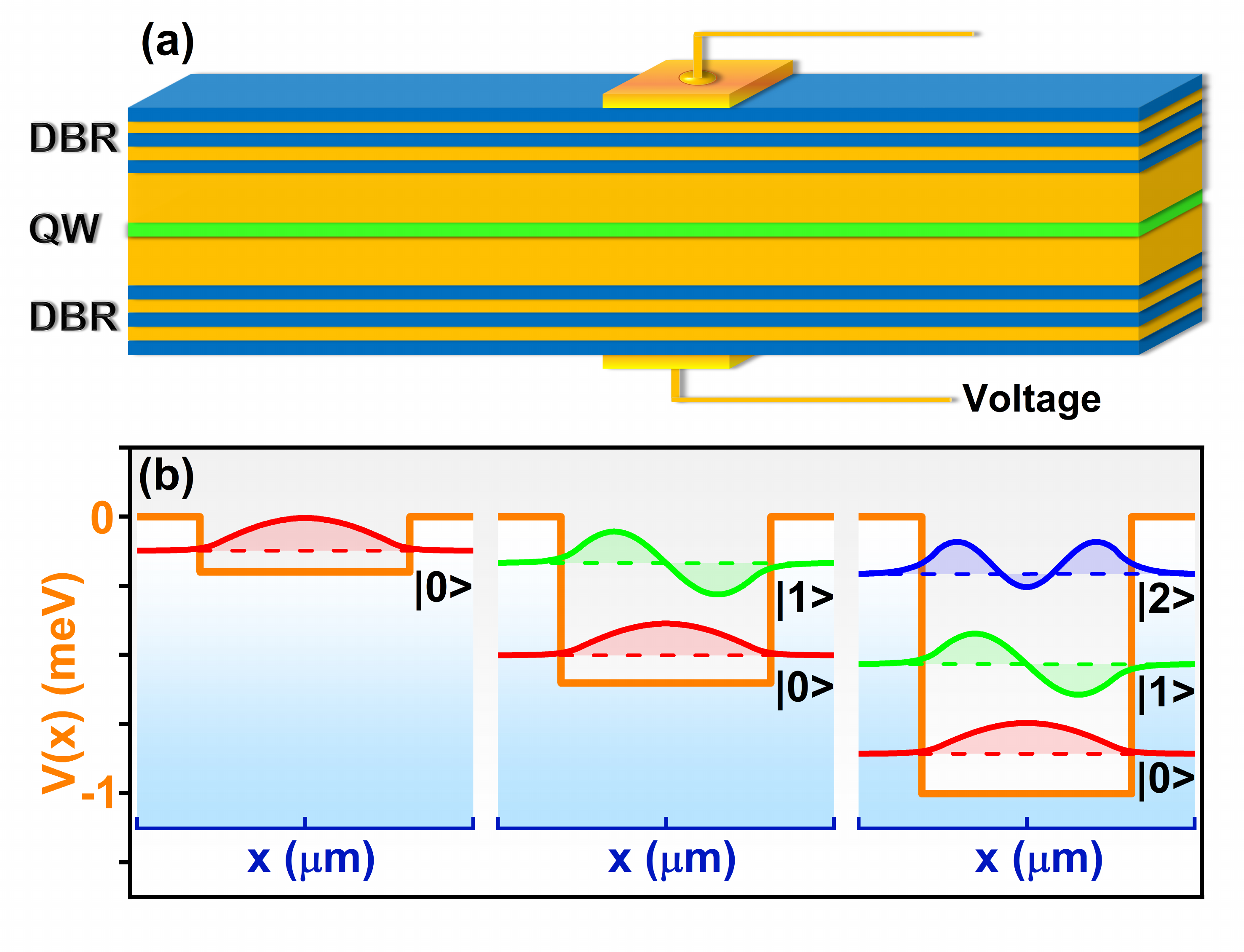}
	\caption{\footnotesize(Color online) (a) sketch of the quasi-1D semiconductor microcavity. The metallic mesa in the central area creates the centrosymmetric finite square potential well with tunable amplitude $V_0$, which can be tuned by applying an electric field. A homogeneous \textit{cw} light pumps the semiconductor microcavity to generate polariton BEC. (b) Quantized energy states in the potential well with potential depth $V_0$. }
	\label{fig:sketch}
\end{figure}

On the other hand, the Kerr nonlinearity has been used to generate intensity squeezing that can surpass the quantum noise limit \cite{Pan_2020,2020prx_Buisson}.
As an effective Kerr medium, the refractive index of the polariton condensate can be controlled by the light intensity \cite{2010prb_Bramati,2020prb_Andreev,2000prb_Yamamoto}, holding the promise to realize a squeezed light source. As such, it has been shown recently the intensity squeezing of polariton condensate at the turning point of the bistability \cite{polariton squeezing_2014}, the periodic squeezing in a Josephson junction \cite{polariton squeezing_2017}, and polariton blockade  \cite{2006prb_blockade_Carusotto,2019natmat_blockade_volz,polariton blockade_2019} in a fiber cavity. However, the squeezing of polariton dark solitons is mainly limited by two obstacles, i.e., the limited stability \cite{2014Kivshar_solitonstability} and the excess noise characterized by a continuum of microcavity polariton modes. The latter is prone to couple with lattice phonons and to reduce the quantum effect, which could be surpassed via the geometric confinement (in the strong localization limit) \cite{Bamba_2010}.  The  challenge is to realize polariton dark soliton squeezing by overcoming the two limitations simultaneously. 

In this work, we study the intensity squeezing of polariton dark solitons near the turning point of bistability, which are stabilized at quantized energy states of the geometric confining potential. Our setting is a semiconductor microcavity pumped nonresonantly with a homogeneous \textit{cw} light beam and patterned with a central potential [see Fig.~\ref{fig:sketch}(a)], leading to localized, quantized energy states. The interference of the discrete states can form polariton dark solitons of even and odd parities, which are  stabilized by the strong confinement of the potential well. We show that the two types of dark solitons exhibit bistable when adiabatically scanning the potential depth up and down at a constant rate. The phase diagram as a function of the nonlinear interaction is obtained. With the help of Feshbach resonance \cite{2019prl_Feshbach,2014nphy_Deveaud}, the squeezing of polariton dark solitons is achieved near the turning point of the bistability curve. Our study might pave a new route to develop low-power squeezed light sources in solid-state devices with the controllable microcavity polaritons.

\textit{\textbf{Model.}---} In our setting, a metallic mesa is embedded in the central area of a quasi-1D planar semiconductor microcavity [see Fig.~\ref{fig:sketch}(a)] to fabricate a centrosymmetric finite potential well, 
\begin{equation}
V(x)=\begin{cases}
\ V_{0}, ~ |x| \le 1 \mu m \\
\ ~0 ,    ~~  |x| > 1\mu m
\end{cases}
\end{equation}
where the depth $V_0$ is tunable by modulating the electric field applied on the mesa \cite{2013hofling_electry,2020Masumoto}.  States in the potential well are quantized, while states out of the potential remain continuous. The deeper the depth $V_0$, the more the quantized states are, as depicted in Fig.~\ref{fig:sketch}(b). A homogeneous \textit{cw} optical field excites non-resonantly polariton particles to form two domains of BEC in real space, separated by the potential well. 

Dynamics of the mean value $\psi$ of the polariton field $\hat{\Psi}$ is governed by an open Gross-Pitaevskii (GP) equation  coupled to the rate equation for the exciton reservoir density $\nR$ \cite{Yamamoto_2012}:
\begin{align}\label{mean1}
	i\hbar \partial \psi 
	=&[\frac{\hbar^2}{2 m} (iD_0 \nR -1)\nabla^2 + V(x) + \gr \nR +  \gc |\psi|^2   \nonumber\\
	&+ \frac{i\hbar}{2}(\R \nR- \gammac)] \psi \partial t +i\hbar dW,\nonumber \\
	\frac{\partial \nR}{\partial t}=&P_{0}-\gammar \nR -\R \nR  |\psi|^2,
\end{align}
where $m=5 \times 10^{-5}m_{e}$ (${m_{e}}$ to be electron mass) is the effective mass of polaritons. $D_{0}=10^{-3}\mu$m and  $\R$ is the energy relaxation constant, and stimulated scattering factor. $\gammac=(50$ps$)^{-1}$ and $\gammar=2\gammac$ are the loss rate of the polariton and exciton reservoirs due to the finite lifetime, respectively. The polaritons are subject to an effective nonlinear Kerr effect originated from the repulsive polariton-polariton interaction with coefficient $gc$, which is controlled by Feshbach resonance~\cite{2019prl_Feshbach,2014nphy_Deveaud}. Coefficient $g_R^{\rm 1D}=2gc$ gives the effective repulsive Coulomb interaction between polaritons and exciton reservoir. $P_0$ is a \textit{cw} optical field tuned well above the polariton resonance. $dW$ is a complex stochastic term that describes the quantum fluctuation. In the truncated Wigner approximation \cite{Michiel_2009}, the correlations of $dW$ are given by 
\begin{align} 
	\langle dW(\mathbf{r},t)dW(\mathbf{r}^\prime,t) \rangle &=0, \ \  \ \langle dW(\mathbf{r},t)dW(\mathbf{r},t^\prime) \rangle =0, \nonumber \\
	\langle dW(\mathbf{r},t)dW^*(\mathbf{r}^\prime,t^\prime)\rangle&=\frac{dt}{2 dxdy}(R n_R+ \gamma_c) \delta_{\mathbf{r},\mathbf{r}^\prime}\delta_{t,t^\prime}, \nonumber
\end{align}
with $\delta_{\mathbf{r},\mathbf{r}'}$ and $\delta_{t,t^\prime}$ to be Kronecker delta function.

\begin{figure} [!htb]
	\centering
	\includegraphics[width=0.9\linewidth]{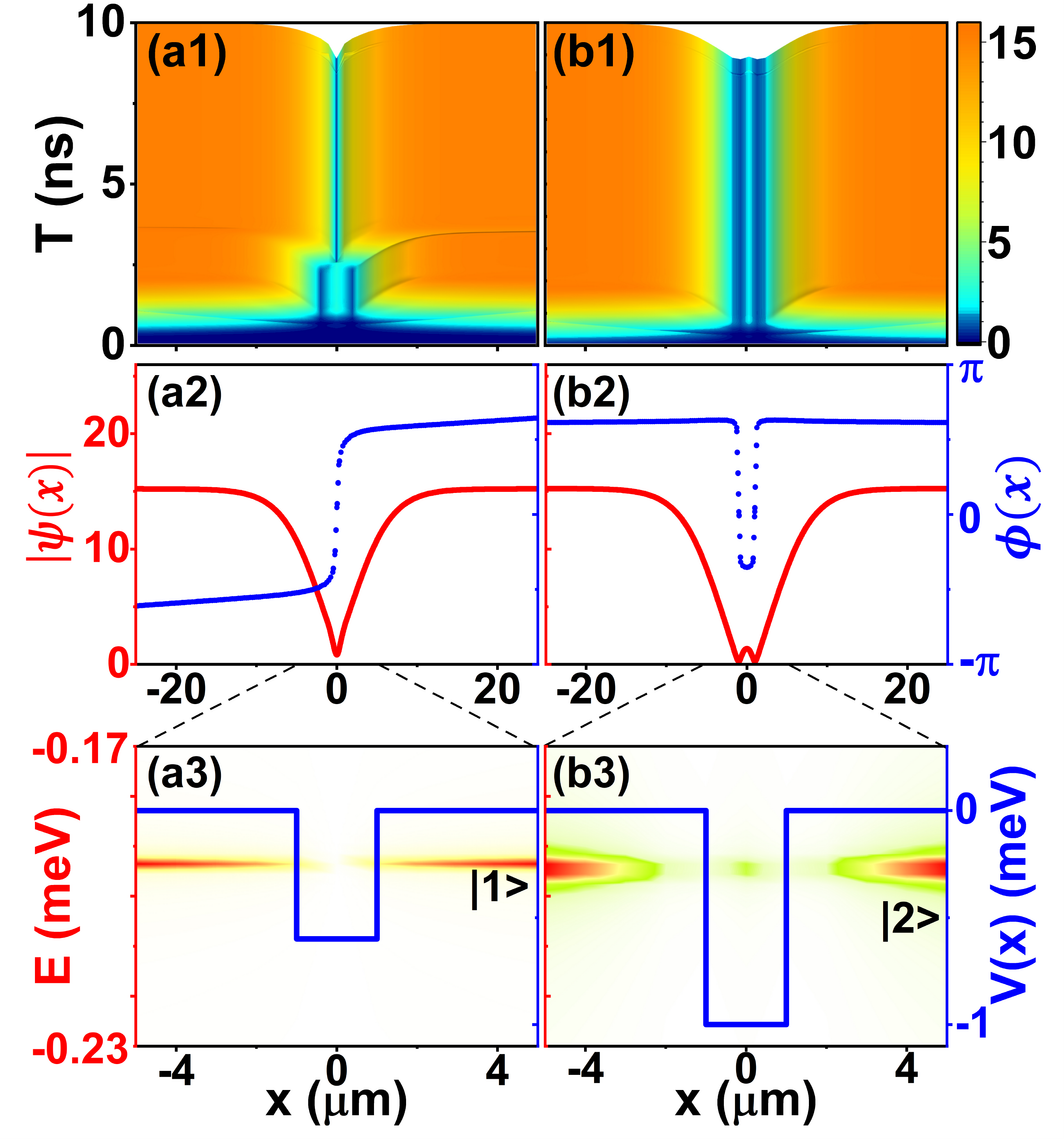}
	\caption{\footnotesize(Color online) (a) Single dark soliton and (b) a pair of sink-type dark solitons. Both solitons are localized in the potential well region. Top row shows the dynamical evolution of polariton density, and the formation of the polarton dark solitons.  Middle row shows polariton amplitude $|\psi(x)|$ and phase $\phi(x)$ at t=10ns, which clearly shows the formation of the dark soliton. Bottom row gives the corresponding spectra of these states in the potential well. Other parameters are $V_0=-0.6$meV in (a) and $V_0=-1.0$meV in (b), while the pumping power $P_0=2.3P_{th}$ and $gc=0.4787 \mu$eV $\mu$m.}
	\label{fig:solitons}
\end{figure}

\textit{\textbf{Polariton dark solitons.}---} 
The quasi-1D polariton condensate is generated when the pump field is above the threshold $P_{0}>P_{th}=\gammac \gammar/R^{1D}$ \cite{2010Bloch_BEC}. Without the potential well ($V_0=0$), polariton density distributes evenly in space with $n_{C}^{0}=(P_{0}-P_{th})/\gammac$ in the steady state. With the centrosymmetric potential well (Fig.~\ref{fig:sketch}), two domains of polariton BEC appear in the extended space across the potential well [see Fig.~\ref{fig:solitons}]. Away from the potential well, the polariton density ($\approx n_{C}^{0}$) is nearly homogeneous. In the potential region, both the polariton density and phase $\phi$ vary dramatically. As the velocity $v\propto\nabla_{x} \phi$, one notes that the polariton particles in these two domains accelerate towards the center of the potential, leading to counter-propagating waves. Due jointly to the high-velocity propagation and phonon-assisted relaxation, polaritons with opposite wave vectors collide in the potential center, and induce a sink-type interference. Depending on whether the interference is constructive or destructive, $\psi$ exhibits \textit{odd}  or \textit{odd} parity [see Fig~\ref{fig:solitons}(a2), Fig.~\ref{fig:solitons}(b2), and discussion below]. 

Dark solitons are stabilized by the nonlinear interaction [see Figs.~\ref{fig:solitons}(a1) and (b1)]. As low-energy excitations, the dark soliton is additionally affected by the pumping field power and potential depth. To illustrate this dependence, we first consider a scenario with fixed $P_{0}=2.3P_{th}$ and $gc=0.4787 \mu$eV $\mu$m. When  -0.8meV$\le V_{0} \le$ -0.5meV, a dark soliton with \textit{odd} parity forms, depicted in Fig.~\ref{fig:solitons}(a). Deepening the potential depth down to -1.6mev$\le V_{0} \le -1.0$meV, a dark soliton of \textit{even} parity is found, originated from the constructive interference between the counter-propagating waves. Note that the two groups of dark solitons might be regarded as complex modes of the bound states $\ket{1}$ and $\ket{2}$. When projected to the bound states, the odd (even) parity soliton maintains the maximum population in state $|1\rangle|$ ($|2\rangle$). This is in sharp contrast to the typically equilibrium case where polariton particles condense on the lowest energy state. 
As shown in Fig. \ref{fig:solitons}, the destructive and constructive interference result in a neat separation of polariton condensate into two domains by a wavy junction in the potential well. The steep jump of the phase across the junction, together with the corresponding density depletion, demonstrates the spontaneous creation of these two types (\textit{odd} or \textit{even}) of dark solitons acting as an insulating barrier~\cite{Sanvitto_2019}. 

\begin{figure} [!htb]
	\centering
	\includegraphics[width=0.9\linewidth]{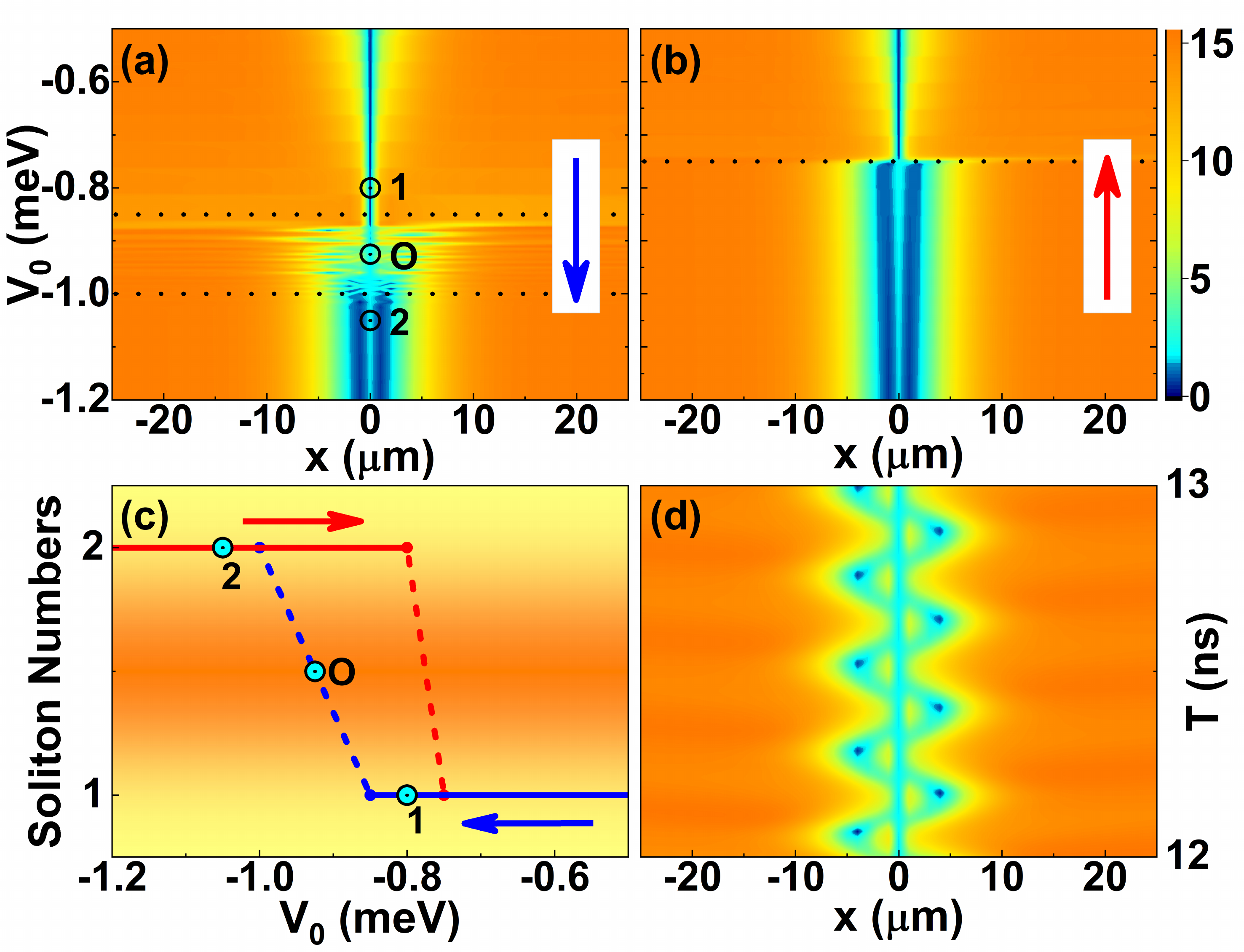}
	\caption{\footnotesize(Color online) (a-b) transition between the \textit{even} and \textit{odd} parity polariton dark solitons. The arrow shows the direction when changing $V_0$. (c) bistability phase diagram of the \textit{even} and \textit{odd} parity  polariton dark solitons by scanning $V_{0}$ with rate $5\times 10^{-5}$meV/ps. One dark soliton is found in the \textit{odd} parity and two dark solitons are found in the \textit{even} parity. The arrows indicate the direction of bistability transitions. (d) snake instability between the two phases at the parameters denoted by O in (a) and (c).} 
	\label{fig:bistability}
\end{figure}

\textit{\textbf{Emergent bistabilities---.}} In the \textit{odd} parity case we obtain a single polariton dark soliton, while the \textit{even} parity could be regarded as two dark solitons, whose phase changes by $-\pi$ and $\pi$, respectively [from left to right in Fig.~\ref{fig:solitons}(b2)]. These two types of dark solitons  can transit from one to the other when the potential depth $V_0$ is adiabatically scanned up and down, as shown in Figs. \ref{fig:bistability}(a) and (b). Uniquely the bistability is controlled entirely by the potential depth $V_{0}$, shown in Fig. \ref{fig:bistability}(c). When $V_0$ is scanned down to the value between the two phases [Fig.~\ref{fig:bistability}(a) and (c)], the polariton condensate is not in a steady state, whose density oscillates periodically [see Fig. \ref{fig:bistability}(d)]. This leads to a snake instability  due to the superposition between the \textit{odd} and \textit{even} parity solitons. Note that  the appearance of the snake instability is directional, i.e. only found in the odd-even transition (decreasing $V_0$). This is largely due to the fact that the sudden birth of the even parity soliton corresponds to a high energy state. The relaxation of this excitation is very slow as a significant fraction of the population in the state $|1\rangle$ and $|2\rangle$. 

\begin{figure} [!htb]
	\centering
	\includegraphics[width=0.9\linewidth]{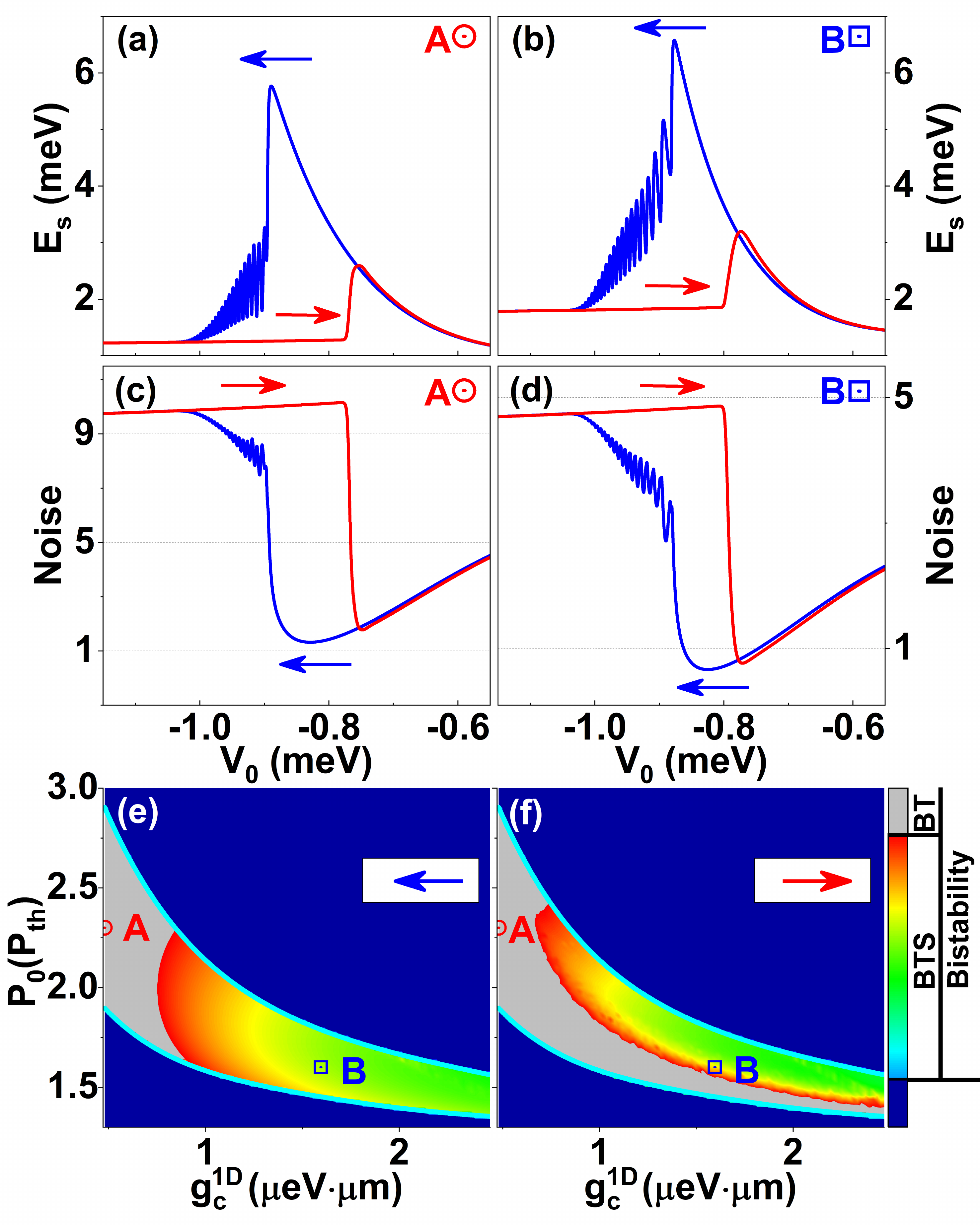}
	\caption{\footnotesize(Color online) (a)-(b) The perturbative energy  and (c)-(d) intensity noise (squeezing parameter) when scanning $V_0$. In (a) the density dependent nonlinear effect dominates, such that squeezing is absent in (c). Increasing $gc$, the nonlinear interaction suppresses the quantum fluctuations, leading to squeezing, shown in (d). The directional dependent phase diagram for both the bistability and squeezing is shown in (e) by decreasing $V_0$, and (f) by increasing $V_0$ across the bistable region, respectively. In (a) and (c) only stability is found [corresponding to point A in (e) and (f) with BT phase], where the parameter $P_0=2.3P_{th}$, $\gc=0.4787 \mu$eV$\cdot\mu$m [the same as in Fig.~\ref{fig:bistability}(c)]. In (b) and (d) both bistability and squeezing are found [corresponding to point B in (e) and (f) with BTS phase]. The related parameters are $P_0=1.6P_{th}$ and $\gc=1.5958 \mu$eV$\cdot\mu$m. In all panels, the arrows indicate the scanning direction of $V_{0}$. }
	\label{fig:squeezing}
\end{figure}

To further illustrate the directional excitation in the bistable transition, we examine the energy of the polariton dark soliton. In the perturbative regime, we assume that the dark soliton wave function $\psi_{d}$ propagates on a background of spatially homogeneous condensate, given by $\psi=\sqrt{n_{C}^{0}} \times \psi_{d}$ and $\nR=\nR^{0} + m_R$. Here $n_{C}^{0}$ and $\nR^{0}$ are spatially homogeneous densities respectively for polaritons and exciton reservoir. $m_R$ is the perturbation for $\nR^{0}$. Using Eq.~(\ref{mean1}), we obtain the energy of the dark soliton \cite{2014prb_elena}:
\begin{align}\label{energy-soliton}
E_s=&\frac{1}{\hbar n_{C}^0}\int [\frac{\hbar^{2}}{2m}(1-iD_0\frac{\gammac}{\R})|\frac{\partial \psi_{d}}{\partial x}|^2 -V(x)|\psi_{d}|^2   \nonumber \\
&+\frac{1}{2}\gc(n_{C}^0-|\psi_{d}|^2)^2]dx.
\end{align}
As shown in Fig.~\ref{fig:squeezing}(a) and (b), the perturbative energy exhibits  different profiles, depending on how the potential depth is varied. It has a similar shape to the bistability transition found in Fig.~\ref{fig:bistability}(c). The oscillation of the energy signifies the snake instability, as seen in the density pattern shown in fig.~\ref{fig:bistability}(d). Note that it is found only when $V_0$ is changed across  the bistable transition.

We want to emphasize that the bistability discussed in this work is realized in an off-resonantly coupled polariton condensate. The bistability is driven by the potential depth $V_0$, and the two phases are stabilized by the nonlinear interaction and spatial confinement. Such bistability is essentially different from the one found in polariton condensates with resonant pumping \cite{2016prl_Bloch}, which is achieved by scanning the phase twist of the pump field with a constant laser power. In both situations, on the other hand, the interference between polariton condensates \cite{2016prl_Lagoudakis} plays key roles in obtaining the different types (parities) of the dark solitons. 

\textit{\textbf{Squeezing of dark solitons---.}} The Kerr-like nonlinear interaction allows to achieve quadrature-squeezed light. Typically the strongest squeezing is found around the turning points of the bistability \cite{squeezing_1992}. We characterize the squeezing of intensity noise through parameter $\xi$ \cite{polariton squeezing_2014, polariton squeezing_2017}:
\begin{align}\label{bvariance}
\xi=\frac{\langle\Delta N\rangle^2}{\langle\hat{\Psi}^{\dagger} \hat{\Psi}\rangle}
\approx \frac{2\langle |\psi|^2 \delta \hat{\psi}^{\dagger} \delta \hat{\psi}\rangle+2\langle Re[\psi^{*2}\delta \hat{\psi} \delta \hat{\psi}]\rangle}{\langle|\psi|^2 + \delta \hat{\psi}^{\dagger} \delta \hat{\psi}\rangle}   
\end{align}
where $\delta \hat{\psi}=\hat{\Psi}-\psi$ and $\delta \hat{\psi}^{\dagger}=\hat{\Psi}^{\dagger}-\psi^{*}$ represent the fluctuation (noise) operator fulfilling $\langle \delta \hat{\psi} \rangle=\langle \delta \hat{\psi}^{\dagger} \rangle=0$, with $\hat{\Psi}$ and $\hat{\Psi}^{\dagger}$ to be the annihilation and creative operators of polariton field. The correlation of fluctuation operators satisfies the following relation,
\begin{align}
	\langle\delta \hat{\psi} \delta \hat{\psi}\rangle=&\gammac M_{11}M_{12}(\langle\hat{\Psi}_{in}^{\dagger} \hat{\Psi}_{in}\rangle+\langle\hat{\Psi}_{in}\hat{\Psi}_{in}^{\dagger}\rangle) \nonumber \\
	\langle \delta \hat{\psi}^{\dagger}\delta \hat{\psi}\rangle=&\gammac M_{21}M_{12}\langle\hat{\Psi}_{in}\hat{\Psi}_{in}^{\dagger}\rangle +\gammac M_{22}M_{11}\langle \hat{\Psi}_{in}^{\dagger} \hat{\Psi}_{in}\rangle \nonumber
\end{align}
where $M_{11}=M_{22}^{*}=\frac{\hbar(i + D_0 \nR )}{2m} \nabla^2 -i\gr \nR - 2i\gc |\psi|^2+\frac{1}{2}(\R \nR -\gamma_c)-i V(x)$ and $M_{12}=M_{21}^{*}=-i \gc \psi^2 $.
$\hat{\Psi}_{in}^{(\dagger)}$ is the input annihilation (correlation) operator of polaritons, which has the form of $\langle\hat{\Psi}_{in}^{\dagger}(t) \hat{\Psi}_{in}(t')\rangle=\sigma P_{0} \delta(t-t')$ and $\langle\hat{\Psi}_{in}(t) \hat{\Psi}_{in}(t')\rangle=0$ ($\sigma$ is a probability constant) \cite{squeezing_1992}. The squeezing parameter determines statistical properties of fluctuations, which is a Poissonian when $\xi=1$, super-Poissonian (bunched) when $\xi>1$, and squeezed (antibunched) when $\xi<1$, correspondingly. 

Both the nonlinear interaction coefficient $\gc$ and pump power plays crucial roles in realizing squeezing ($\xi<1$). To demonstrate this,  we first examine the squeezing parameter corresponding to the situation discussed in Fig.~\ref{fig:squeezing}(a). As shown in Fig.~\ref{fig:squeezing}(c), the intensity noise becomes weaker at the turning points of bistability, where $\xi>1$ is relatively smaller (larger) along the odd-even (even-odd) transition. However irrelevant to the direction, squeezing is not found in either case. Both the topological excitation of polariton dark solitons and the intensity squeezing require the strong nonlinear interaction. We note that the  polariton dark solitons are classical excitations that depend strongly on the pump field. The stronger the pump power, the higher the polariton density. Though this could generate stronger density-dependent nonlinear effects (e.g., stabilizing the dark soliton), quantum fluctuations also increases significantly at the same time [Fig.~\ref{fig:squeezing}(c)], such that squeezing of the noise is not seen.  

Thanks to the Feshbach resonance \cite{2019prl_Feshbach,2014nphy_Deveaud}, we can maintain strong nonlinearities (via tuning $\gc$) even at low polariton densities. In this case, bistability is still clearly visible, as shown in Fig.~\ref{fig:squeezing}(b). Squeezing, however, is drastically improved, as can be seen in Fig.~\ref{fig:squeezing}(d). It is found that $\xi$ becomes smaller round the bistable transition region. More importantly,  squeezing, i.e. $\xi<1$, is achieved around the turning point of the bistability, which is independent of the direction when changing $V_0$. The squeezing strength, however, depending on the direction. Our numerical simulations show that $\xi=0.667$ at the turning point of the bistability when decreasing $V_0$. In contrast, a slightly weaker squeezing with $\xi=0.768$ is obtained when increasing $V_0$.

Through large scale numerical simulations, we obtain phase diagrams to capture both bistability and squeezing of polariton dark solitons, shown in Fig.~\ref{fig:squeezing}(e) and (f). To obtain the phase diagram, dynamics of the system is solved by scanning $V_0$ downwards [Fig.~\ref{fig:squeezing}(e)] and upwards [Fig.~\ref{fig:squeezing}(f)] with a given set of $\gc$ and $P_0$. By examining stability behaviors and squeezing parameters, three distinctive phases, one without bistability (blue region), one with only bistability (BT phase, grey region), and one with both bistability and squeezing  (BTS phase, colors other than blue or grey), are identified. For a given $\gc$, the BT phase forms when the pump field is strong enough, such that the density (Kerr) nonlinearity can stabilize the two types of solitons. Squeezing, on the other hand, occurs when the nonlinear interaction coefficient $\gc$ is large enough while the polariton density is relatively low.  Hence the BT phase occupies larger areas than the BTS phase. Such feature is independent of the changing direction of $V_0$, which is consistent with the example shown in Fig.~\ref{fig:squeezing}(c) and (d). In the phase diagram,  the strongest squeezing is $\xi=0.516$, which is obtained at $\gc=2.473 \mu$eV$\cdot\mu$m and $P_0=1.56P_{th}$ when $V_0$ is decreasing.

\textit{\textbf{Conclusions---.}} We have shown that the odd and even parity dark solitons can be generated in a controlled fashion in a nonresonantly pumped semiconductor microcavity containing a central potential well. The solitons are formed due to the strong geometric confinement and nonlinear Kerr like interaction. By varying the potential depth, a bistable phase of the two dark solitons is found. We have revealed that strong  intensity squeezing can be achieved at the turning points of the bistability. Through large scale numerical simulations, we have demonstrated that the level of the intensity squeezing can be controlled by changing either $P_0$ or $\gc$, where the latter is achieved via the Feshbach resonance~\cite{2019prl_Feshbach,2014nphy_Deveaud}. Our study might open a way to create ultra-low threshold, on-chip integrated topological excitations and squeezed light sources  for quantum information-processing applications.

\begin{acknowledgments}
This work is supported by National Natural Science Foundation of China (Grant No. 12074147).
W. Li acknowledges support from the UKIERI-UGC Thematic Partnership (IND/CONT/G/16-17/73), and the Royal Society through the International Exchanges Cost Share award No. IEC$\backslash$NSFC$\backslash$181078. 
\end{acknowledgments}
~~~~~~~~~~~~~~~~~~~~~~~~~~~~~~~~~~~~~~~~~~~~~

~~~~~~~~~~~~~~~~~~~~~~~~~~~~~~~~~~~~~~~~~~~~~~~~~~~~~~~~~~~~~~~~~~~~~~~~
\clearpage
\begin{widetext}
	\section{Supplementary to "Controllable Bistability and Squeezing of Confined Polariton Dark Solitons"}
	
	\subsection{Soliton energy}
	In the perturbative regime, we assume that the dark soliton wave function $\psi_{d}$ propagates on a background of spatially homogeneous condensate, given by $\psi=\sqrt{n_{C}^{0}} \times \psi_{d}$ and $\nR=\nR^{0} + m_R$. Here $n_{C}^{0}$ and $\nR^{0}$ are spatially homogeneous densities respectively for polaritons and exciton reservoir. $m_R$ is the perturbation for $\nR^{0}$. Substituting these definitions into Eq.~(\ref{mean1}) in the main text, we obtain the following dynamical equations:
	\begin{align}\label{pertur-eqns}
		&i \frac{\partial \psi_d}{\partial t}=-\frac{\hbar}{2m}(1-D_0\frac{\gammac}{R}i)\frac{\partial^2 \psi_d}{\partial x^2}-\gc  n_C^0(1-|\psi_d|^2)\psi_d 
		+V(x)\psi_d + (i D_0 \frac{\partial^2 \psi_d}{\partial x^2}+\frac{i}{2}\R+\gr)m_R \psi_d   \nonumber \\
		&\frac{\partial m_R}{\partial t}=(P_0-P_{th})(1-|\psi_d|^2)-\gammar (1-|\psi-d|^2)m_R -\gammar \frac{P_0}{P_{th}}|\psi-d|^2 m_R
	\end{align}
	According to the Lagrange variational method:
	\begin{align}\label{Lagrange}
		\frac{\delta L}{\delta f}=\frac{\partial}{\partial z}\frac{\partial L}{\partial_z f}+\frac{\partial}{\partial \tau}\frac{\partial L}{\partial_\tau f}-\frac{\partial L}{\partial f} =\int (R \frac{\partial \psi^{*}_d}{\partial f}+R^{*}\frac{\partial \psi_d}{\partial f})dx 
	\end{align}
	The Lagrangian could be constructed from Eq.(\ref{pertur-eqns}):
	\begin{align}\label{Lagrangian}
		L=\frac{i}{2}(\psi_d \frac{\partial \psi^{*}_d}{\partial t}+\psi^{*}_d\frac{\partial \psi_d}{\partial t})+\frac{\hbar}{2m}(1-D_0\frac{\gammac}{R}i)|\frac{\partial \psi_d}{\partial x}|^2 +\frac{1}{2}\gc n_C^0(1-|\psi_d|^2)^2-V(x)|\psi_d|^2
	\end{align}
	So, the energy of the dark soliton is expressed as:
	\begin{align}\label{energy-soliton}
		E_s=\int dx [\frac{\hbar}{2m}(1-iD_0\frac{\gammac}{R})|\frac{\partial \psi_d}{\partial x}|^2 +\frac{1}{2}\gc n_C^0(1-|\psi_d|^2)^2-V(x)|\psi_d|^2]  
	\end{align}
	
	\subsection{Intensity noise}
	Dynamics of the polariton field is modelled by the non-Hermine Hamiltonian of polariton condensate in the vicinity of the ground state, as can be described as \cite{polariton squeezing_2017}:
	\begin{align}\label{hamiltonian1}
		\hat{H}=& \int dx [\hat{\Psi}^{\dagger}(x)(- \frac{\hbar^2}{2 m_{eff}}(1-i D_0 \nR) \nabla^2  + \hbar V(x))\hat{\Psi}(x) + \hbar \gc \hat{\Psi}^{\dagger}(x) \hat{\Psi}^{\dagger}(x)\hat{\Psi}(x)\hat{\Psi}(x)+ \hbar\gr \nR \hat{\Psi}^{\dagger}(x)\hat{\Psi}(x) \nonumber \\
		&+\frac{i\hbar}{2}(\R \nR -\gammac)\hat{\Psi}(x) ]
	\end{align}
	where $\hat{\Psi}(x)$ and $\hat{\Psi}^{\dagger}(x)$ are, respectively, the annihilation and creative operators of a polariton field on the lower-polariton branch. Governed by the Heisenberg equation $i\hbar\frac{\partial{\hat{\Psi}(x)}}{\partial{t}}=-[\hat{H},\hat{\Psi}(x)]$, the Langevin equation for the polariton field is derived as follows: 
	\begin{align}\label{Langevin}
		i\hbar\frac{\partial \hat{\Psi}(x)}{\partial t}
		=&[- \frac{\hbar^2}{2 m}(1-i D_0 n_R) \nabla^2 + \hbar V(x)]\hat{\Psi}(x) + \hbar\gc  \hat{\Psi}^{\dagger}(x) \hat{\Psi}(x)\hat{\Psi}(x)+\hbar \gr \nR \hat{\Psi}(x) + \frac{i\hbar}{2}(\R \nR -\gammac)\hat{\Psi}(x) \nonumber \\
		& -i\hbar\sqrt{\gammac}\hat{\Psi}_{in}(x)
	\end{align}
	here  $\hat{\Psi}_{in}(x)$ is the input operator from the polariton field and assumed to have a zero average $\langle \hat{\Psi}_{in}(x) \rangle=0$ (incoherent input). 
	
	To allow for the large populations involved in polariton condensate, we expand the polariton field operator as $\hat{\Psi}(x)=(\psi+\delta \hat{\psi})$, where $\psi=\langle \hat{\Psi}(x) \rangle$ is the coherent mean field component, $\delta \hat{\psi}$ represents the fluctuation (noise) operator fulfilling $\langle \delta \hat{\psi} \rangle=0$. Substituting $\langle \hat{\Psi}(x) \rangle=\psi$ into the Langevin equation, we first calculate the mean field component of polariton condensate through the Gross-Pitaevskii equation, coupled to a rate equation for the density of excitation reservoir, which are expressed in Eq.~(\ref{mean1}) in the main text.
	
	Next Let's derive the expression of intensity noise with the consideration of fluctuation operator $\delta \hat{\psi}=\hat{\Psi}-\psi$. Substituting it into the Eq. (\ref{Langevin}), the quantum Langevin equation with the assumption of linear approximation is expressed as following:
	\begin{align}\label{quantum}
		i\hbar\frac{\partial \delta \hat{\psi}}{\partial t}&=[-\frac{\hbar^2}{2m} (1-iD_0\nR)\nabla^2+\hbar V(x)+\hbar\gr\nR +2\hbar\gc |\psi|^2 + \frac{i\hbar}{2}(\R \nR -\gammac)]\delta \hat{\psi} +\hbar\gc |\psi|^2 \delta \hat{\psi}^{\dagger} -i\hbar\sqrt{\gammac}\hat{\Psi}_{in}
	\end{align}
	According to Eq.~(\ref{quantum}), we define the following matrix:
	\begin{equation}
	A=\left(
	\begin{array}{c}
	\delta \hat{\psi}~  \\
	\delta \hat{\psi}^{\dagger}~ \\
	\end{array}
	\right)
	~~~~~~
	B=\left(
	\begin{array}{c}
	\sqrt{\gammac} ~ \hat{\Psi}^{in} \\
	\sqrt{\gammac} ~ \hat{\Psi}^{\dagger in} \\
	\end{array}
	\right)  
	~~~~~~
	N=\left(
	\begin{array}{cccccc}
	(i + D_0 \nR )f_{0}+ f_{1} - i f_{2} 	&-i \gc ~\psi^2 	   			\\
	\\
	i \gc ~(\psi^{*})^2						&(-i + D_0 \nR )f_{0}+ f_{1} + i f_{2}     	\\
	\\
	\end{array}
	\right)
	\end{equation}
	with $f_{0}= \frac{\hbar}{2m} \nabla^2 $, $f_{1}=\frac{1}{2}(\R \nR -\gamma_c)$, $f_{2}= \gr \nR + 2\gc |\psi|^2+V(x) $. {\Huge }
	Then Eq.~(\ref{quantum})can be rewirtten as:
	\begin{align}\label{partial}
		\frac{\partial}{\partial t}A=N A-B
	\end{align}
	Set $\frac{\partial}{\partial t}A=0$, we have $A=N^{-1} B=M B$ as follows:
	\begin{equation}\label{stable}
	\left(
	\begin{array}{c}
	\delta \hat{\psi}~  \\
	\delta \hat{\psi}^{\dagger}~ \\
	\end{array}
	\right)
	=\left(
	\begin{array}{cccc}
	M_{11} & M_{12}\\
	M_{21} & M_{22}\\
	\end{array}
	\right)
	\left(
	\begin{array}{c}
	\sqrt{\gammac} ~ \hat{\Psi}^{in} \\
	\sqrt{\gammac} ~ \hat{\Psi}^{\dagger in} \\
	\end{array}
	\right)  
	\end{equation} 
	with $M_{11}=M_{22}^{*}=\frac{(-i + D_0 \nR )f_0   +f_1 + i f_2}{a}$, $M_{12}=M_{21}^{*}=\frac{i \gc ~\psi^2}{a}$ and $a=(f_2-f_0)^2+(f_1+D_0 \nR f_0)^2-(\gc |\psi|^2)^2$. The correlation of fluctuation operator can be derived as:
	\begin{align}\label{bcorrelation}
		\langle \delta \hat{\psi}^{\dagger}\delta \hat{\psi} \rangle=&\gammac M_{21}M_{12}\langle \hat{\Psi}_{in}\hat{\Psi}_{in}^{\dagger} \rangle+\gammac M_{22}M_{11}\langle \hat{\Psi}_{in}^{\dagger} \hat{\Psi}_{in} \rangle  \nonumber \\
		\langle \delta \hat{\psi} \delta \hat{\psi} \rangle=&\gammac M_{11}M_{12}(\langle \hat{\Psi}_{in}^{\dagger} \hat{\Psi}_{in} \rangle+\langle \hat{\Psi}_{in}\hat{\Psi}_{in}^{\dagger} \rangle )
	\end{align}
	where $M_{11}=M_{22}^{*}=\frac{\hbar(i + D_0 \nR )}{2m} \nabla^2 -i\gr \nR - 2i\gc |\psi|^2+\frac{1}{2}(\R \nR -\gamma_c)-i V(x)$ and $M_{12}=M_{21}^{*}=-i \gc ~\psi^2 $.
	Assume that the correlation of input operator of polariton has the form of $\langle \hat{\Psi}_{in}^{\dagger}(t) \hat{\Psi}_{in}(t') \rangle=\sigma P_{0} \delta(t-t')$ and $\langle \hat{\Psi}_{in}(t) \hat{\Psi}_{in}(t') \rangle=0$ ($\sigma$ is a probability constant). With this assumption and the input-output relation, we could have the results for the intensity noise according to the following expression:
	\begin{align}\label{bvariance}
		\frac{\langle\Delta N\rangle^2}{\langle\hat{\Psi}^{\dagger} \hat{\Psi}\rangle}&=\frac{\langle\hat{\Psi}^{\dagger} \hat{\Psi}^{\dagger} \hat{\Psi} \hat{\Psi}\rangle-\langle\hat{\Psi}^{\dagger} \hat{\Psi}\rangle^2 }{\langle\hat{\Psi}^{\dagger} \hat{\Psi}\rangle}  =\frac{\langle|\psi|^{4}+4|\psi|^2 \delta \hat{\psi}^{\dagger} \delta \hat{\psi}+2Re[\psi^{*2}\delta \hat{\psi} \delta \hat{\psi}]\rangle-\langle|\psi|^2 + \delta \hat{\psi}^{\dagger} \delta \hat{\psi}\rangle^{2}}{\langle|\psi|^2 + \delta \hat{\psi}^{\dagger} \delta \hat{\psi}\rangle} \nonumber \\
		&\approx \frac{2\langle |\psi|^2 \delta \hat{\psi}^{\dagger} \delta \hat{\psi}+Re[\psi^{*2}\delta \hat{\psi} \delta \hat{\psi}]\rangle}{\langle|\psi|^2 + \delta \hat{\psi}^{\dagger} \delta \hat{\psi}\rangle} 
	\end{align}
	Generally, the intensity noise characterizes the statics of light: it is Poissonian when equals to $1$, super-Poissonian(bunched) when larger than $1$, and quantum (antibunched) when smaller than $1$.
\end{widetext}
\end{document}